\documentclass[aps,prd,groupedaddress,nofootinbib,superscriptaddress,twocolumn]{revtex4}
\usepackage{amsfonts}
\usepackage{amssymb}
\usepackage{amsmath,color}
\usepackage{epsfig}
\usepackage{graphicx,epsf,epsfig}
\usepackage{bbm}
\def\slc#1{\setbox0=\hbox{$#1$}           % set a box for #1
    \dimen0=\wd0                                 % and get its size
    \setbox1=\hbox{/} \dimen1=\wd1               % get size of /
    \ifdim\dimen0>\dimen1                        % #1 is bigger
       \rlap{\hbox to \dimen0{\hfil/\hfil}}      % so center / in box
       #1                                        % and print #1
    \else                                        % / is bigger
       \rlap{\hbox to \dimen1{\hfil$#1$\hfil}}   % so center #1
       /                                         % and print /
    \fi}

\newcommand{\dms}{\mbox{$\Delta m^2_{\odot}$}}
\newcommand{\dma}{\mbox{$\Delta m^2_{\rm A}$}}
\def\gs{\mathrel{
   \rlap{\raise 0.511ex \hbox{$>$}}{\lower 0.511ex \hbox{$\sim$}}}}
\def\ls{\mathrel{
   \rlap{\raise 0.511ex \hbox{$<$}}{\lower 0.511ex \hbox{$\sim$}}}}

\begin{document}
%----------------------------------------------------------------------------------
\title{Extended Empirical Fermion Mass Relation}
%%----------------------------------------------------------------------------------
%\date{\today}
%%----------------------------------------------------------------------------------

\author{Werner Rodejohann}
\email{werner.rodejohann@mpi-hd.mpg.de}

\affiliation{Max-Planck-Institut f{\"u}r Kernphysik, Postfach
103980, 69029 Heidelberg, Germany}

\author{He Zhang}
\email{he.zhang@mpi-hd.mpg.de}

\affiliation{Max-Planck-Institut f{\"u}r Kernphysik, Postfach
103980, 69029 Heidelberg, Germany}
%----------------------------------------------------------------------------------
%\pacs{12.10.-g, 12.60.Jv, 14.60.Pq, 12.15.Ff}
%----------------------------------------------------------------------------------
\begin{abstract}

It is known that the charged lepton masses obey to high precision an
interesting empirical relation (Koide relation). In turn, the light
neutrino masses cannot obey such a relation. We note that if
neutrinos acquire their mass via the seesaw mechanism, the empirical
mass relation could hold for the masses in the Dirac and/or heavy
Majorana mass matrix. Examples for the phenomenological consequences
are provided. We furthermore modify the mass relation for light
neutrino masses including their Majorana phases, and show that it
can be fulfilled in this case as well, with interesting predictions
for neutrinoless double beta decay. Finally, we remark that while
the relation  does not hold for the up- and down-quarks, it may be
valid for the $u, d, s$ quarks, and for the $c, b, t$ quarks.

\end{abstract}
\maketitle
%%%%%%%%%%%%%%%%%%%%%%%%%%%%%%%%%%%%%%%%%%%

\section{Introduction} \label{sec:intro}

The experimentally measured charged-lepton masses reveal that they
obey the following empirical mass relation
(Koide relation)~\cite{Koide:1982si,Koide:1983qe}
\begin{eqnarray}\label{eq:koide-relation}
K_\ell=\frac{m_e+m_\mu+m_\tau}{\left(\sqrt{m_e}+\sqrt{m_\mu}+\sqrt{m_\tau}\right)^2}
\simeq \frac{2}{3} \,
\end{eqnarray}
with remarkable precision, i.e., the above relation is correct to
${\cal O}(10^{-5})$. A number of authors have tried to understand
Eq.~\eqref{eq:koide-relation} based on possible flavor symmetries
and phenomenological
conjectures~\cite{Koide:1989jq,Foot:1994yn,Koide:1995xk,Koide:1996yf,Rivero:2005vj,Krolikowski:2005kr,Gerard:2005ad,Koide:2008tr,Sumino:2008hu,Sumino:2008hy,Sumino:2009bt,Koide:2010np}.

Motivated by the idea of grand unification, one may wonder whether a
similar mass relation exists for other fermions. Numerical analysis
has shown that neither the up-quark masses, down-quark masses nor the
light neutrino masses could satisfy such an empirical relation, even if the
renormalization group (RG) running effects are taken into
account~\cite{Li:2006et,Xing:2006vk}. Consider for illustration
the charged leptons and ignore the small ratios
$m_e/m_\tau$, $\sqrt{m_e/m_\tau}$ and $m_\mu/m_\tau$ in
Eq.~\eqref{eq:koide-relation}. One approximately obtains
\begin{eqnarray}\label{eq:Kl}
K_\ell \simeq \frac{1}{ 1+2\sqrt{\displaystyle {m_\mu}/{m_\tau}}}
 \, .
\end{eqnarray}
By using the pole masses of charged leptons given in Particle Data
Group~\cite{Nakamura:2010zzi}, one can estimate that
$\sqrt{m_\mu/m_\tau} \simeq 0.25$, which roughly yields $K_\ell =
2/3$. Obviously, if the mass of the lightest particle can be
neglected, the empirical relation relies on the mass ratio between
the two heavier masses. In the extreme case of a highly hierarchical
mass spectrum, in which the ratio of the second heaviest and the
heaviest mass is very small, one has $K \to 1$. This is the case for
the up- and down-quarks, for which the values $K_{\rm up} \simeq
0.89 $ and $K_{\rm down} \simeq 0.75$ apply (both evaluated at $M_Z
= 91.2$ GeV~\cite{Xing:2007fb}). The other extreme limit applies
when the mass spectrum is nearly degenerate, in which case $K \to
1/3$. In this sense, the empirical mass relation
Eq.~\eqref{eq:koide-relation} appears interestingly exactly in the
middle of all its possible values\footnote{For $N$ fermion
generations the range would be between 1 and $1/N$, with a central
value of $(N+1)/2N$.}. One can understand also why light neutrinos
cannot obey the empirical relation. If they are quasi-degenerate,
$K_\nu \simeq \frac 13$, and if they obey a normal hierarchical
spectrum,
\begin{equation}
K_\nu \simeq \frac{1}{1 + 2\sqrt[4]{\dms/\dma}} \ls 0.55 \, ,
\end{equation}
where $\dms$ and $\dma$ are the solar and atmospheric
mass-squared differences.

\begin{figure}[t]
\begin{center}\vspace{0.cm}
\includegraphics[width=7cm,height=5.6cm]{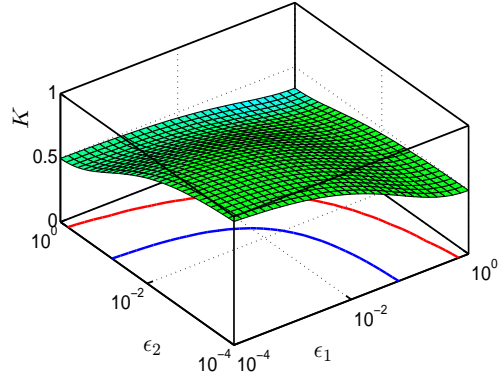}
\vspace{0.cm} \caption{\label{fig:fig0} The allowed range of $K$
with respect to the ratios $\epsilon_1$ and $\epsilon_2$. The red
and blue curves on the $\epsilon_1$-$\epsilon_2$ plane indicate
$K=1/2$ and $K=2/3$, respectively.} \vspace{-0.3cm}
\end{center}
\end{figure}

We now extend the expression in Eq.~\eqref{eq:koide-relation} for
arbitrary three masses, $m_x$, $m_y$ and $m_z$, i.e.,
\begin{eqnarray}\label{eq:K}
K \equiv \frac{m_x + m_y +
m_z}{\left(\sqrt{m_x}+\sqrt{m_y}+\sqrt{m_z}\right)^2} = \frac{1 +
\epsilon_1 +
\epsilon_2}{\left(1+\sqrt{\epsilon_1}+\sqrt{\epsilon_2}\right)^2}
 \, ,
\end{eqnarray}
where $\epsilon_1=m_x/m_z$ and $\epsilon_2 = m_y / m_z$. The allowed
range of $K$ is shown in Fig.~\ref{fig:fig0}.
We have in particular indicated the values of $\epsilon_1$ and
$\epsilon_2$, which give $K = \frac 23$.
One can read from the plot that the minimum value $K=1/3$ appears at
the position $\epsilon_1=\epsilon_2=1$, while the maximum value
$K=1$ is obtained for $\epsilon_1=\epsilon_2=0$. In addition, if
$\epsilon_1$ ($\epsilon_2$) is vanishing, a lower bound $K \geq 1/2$
can be achieved for $\epsilon_2=1$ ($\epsilon_1=1$).

As mentioned above, in the quark sector both up-type and down-type
quark mass spectra are quite hierarchical, and the $K$-value is
generally larger than $2/3$. However, if the quarks are divided in
light and heavy quarks, instead of up- and down-like quarks (i.e.,
according to mass instead of their electric charge or isospin), the
empirical relation could be well satisfied. For example, considering
the pole masses of the heavy quarks at 1$\sigma$~C.L., the ratio
\begin{eqnarray}\label{eq:koide-heavy}
K_{\rm heavy} =
\frac{m_c+m_b+m_t}{\left(\sqrt{m_c}+\sqrt{m_b}+\sqrt{m_t}\right)^2}
\end{eqnarray}
is found to be $0.66<K_{\rm heavy}<0.68$. This is in amazing
agreement with Eq.~\eqref{eq:koide-relation}. Due to the
non-perturbative nature of quantum chromodynamics at low energies,
the pole masses of light quarks are not well defined. Therefore, we
make use of the running light quark masses at the scale
$M_Z=91.2~{\rm GeV}$, and find that a similar relation for light
quarks
\begin{eqnarray}\label{eq:koide-light}
K_{\rm light} =
\frac{m_u+m_d+m_s}{\left(\sqrt{m_u}+\sqrt{m_d}+\sqrt{m_s}\right)^2}
\end{eqnarray}
is located in the interval $0.49<K_{\rm light}<0.65$, indicating a
small deviation from the exact empirical relation at about
1$\sigma$~C.L.

We will focus on the application of the mass relation to the
neutrinos in the rest of the paper. As mentioned above, their mass
spectrum is not hierarchical enough to reproduce the empirical
relation, and $1/3 \ls K_\nu \ls3/5$ typically holds, regardless of
the neutrino mass ordering~\cite{Xing:2006vk}. However, we would
like to note here that if neutrinos are Majorana particles their
mass presumably originates from the seesaw
mechanism~\cite{Minkowski:1977sc,Yanagida:1979as,GellMann:1980vs,Mohapatra:1979ia},
i.e., the light neutrino masses appear as a combination of Dirac and
Majorana mass terms:
\begin{eqnarray}\label{eq:seesaw}
m_\nu =M_{\rm D} M^{-1}_{\rm R} M^T_{\rm D} \, .
\end{eqnarray}
In this case, it is not surprising that the empirical mass relation
does not directly apply to neutrino masses. Instead, it is much more
natural to assume that the relation is fulfilled in $M_{\rm D}$ and/or
$M_{\rm R}$. Therefore, in this paper we will extend the empirical mass relation
to the seesaw framework, and study some of the phenomenological consequences
resulting from this hypothesis. Another crucial aspect of Majorana
neutrinos is the presence of Majorana phases, and we will also modify
the empirical relation taking this into account. It is shown
that the relation can work in this case as well.

The rest of the paper is organized as follows: In
Sec.~\ref{sec:seesaw relation}, we introduce the seesaw extended
mass relation, and describe four typical scenarios, in which very
distinctive predictions on the light and heavy neutrino mass
spectrum can be gained. The phenomenological consequences of these
four scenarios are figured out, and the parameter spaces are
illustrated. In Sec.~\ref{sec:variation}, we generalize the
empirical mass relation by including contributions from CP-violating
phases, and present the constraints on the Majorana phases as well
as the light neutrino masses. In particular, we present the
predictions on the effective mass relevant for neutrinoless double beta decay.
Finally, in Sec.~\ref{sec:summary}, we summarize our work
and state our conclusions.

\section{Mass relation in the seesaw model} \label{sec:seesaw relation}

Without loss of generality, one can always work in a basis in
which the right-handed neutrino Majorana mass matrix is diagonal,
i.e., $M_{\rm R}={\rm diag}(M_1,M_2,M_3)$ with $M_i$ (for $i=1,2,3$)
being the masses of right-handed neutrinos. If lepton mixing stems
entirely from the charged lepton sector, as possible in flavor
symmetry models, the Dirac mass term $M_{\rm D}$ is diagonal, i.e., $M_{\rm D}={\rm
diag}(D_1,D_2,D_3)$. This is a strong assumption, but very
helpful for illustrating the point we wish to make in this note.
We will comment later on the general case.
The light neutrino masses are simply given by
\begin{eqnarray}\label{eq:mass}
m_i = D^2_i/M_i \, .
\end{eqnarray}
We stress again that though the empirical relation is shown to be
not compatible with the light neutrino masses, the exact relation
may exist in the mass matrices $M_{\rm D}$ and/or $M_{\rm R}$. We
will discuss 4 different cases.

\begin{itemize}

\item Case I: The empirical relation exist in $M_{\rm R}$ whereas
$M_{\rm D}$ is an identity matrix, i.e., $D_i = D_0$. In
this limit, the right-handed neutrino masses are proportional to the
inverse of the light neutrino masses. A hierarchical mass
spectrum of right-handed neutrinos can be achieved if one of the
light neutrino masses is extremely small. For example, in the normal
hierarchy case $m_1 \ll m_2 \ll m_3 $, one obtains
\begin{eqnarray}\label{eq:KRnor}
K_{\rm R} \simeq  \frac{1}{1+2\sqrt{m_1/m_2}+2\sqrt{m_1/m_3}}  \, ,
\end{eqnarray}
which gives $K_{\rm R} \simeq 2/3$ if $m_1/m_2 \simeq 0.03$. In the
inverted hierarchy case one has $m_3 \ll m_1 < m_2 $, and
\begin{eqnarray}\label{eq:KRinv}
K_{\rm R} \simeq  \frac{1}{1+4\sqrt{m_3/m_2}} \, ,
\end{eqnarray}
can be expected. The empirical relation requires $m_3/m_2 \simeq
0.016$.

We show in the upper plot of Fig.~\ref{fig:fig1} the dependence of
$K_{\rm R}$ on the lightest neutrino mass in case I. In the
numerical computations, we make use of the values of neutrino
mass-squared differences from a global-fit of current neutrino
oscillation experiment~\cite{Schwetz:2008er}, and allow them to vary
in their $1\sigma$ interval.
\begin{figure}[t]
\begin{center}\vspace{0.1cm}
\includegraphics[width=7cm]{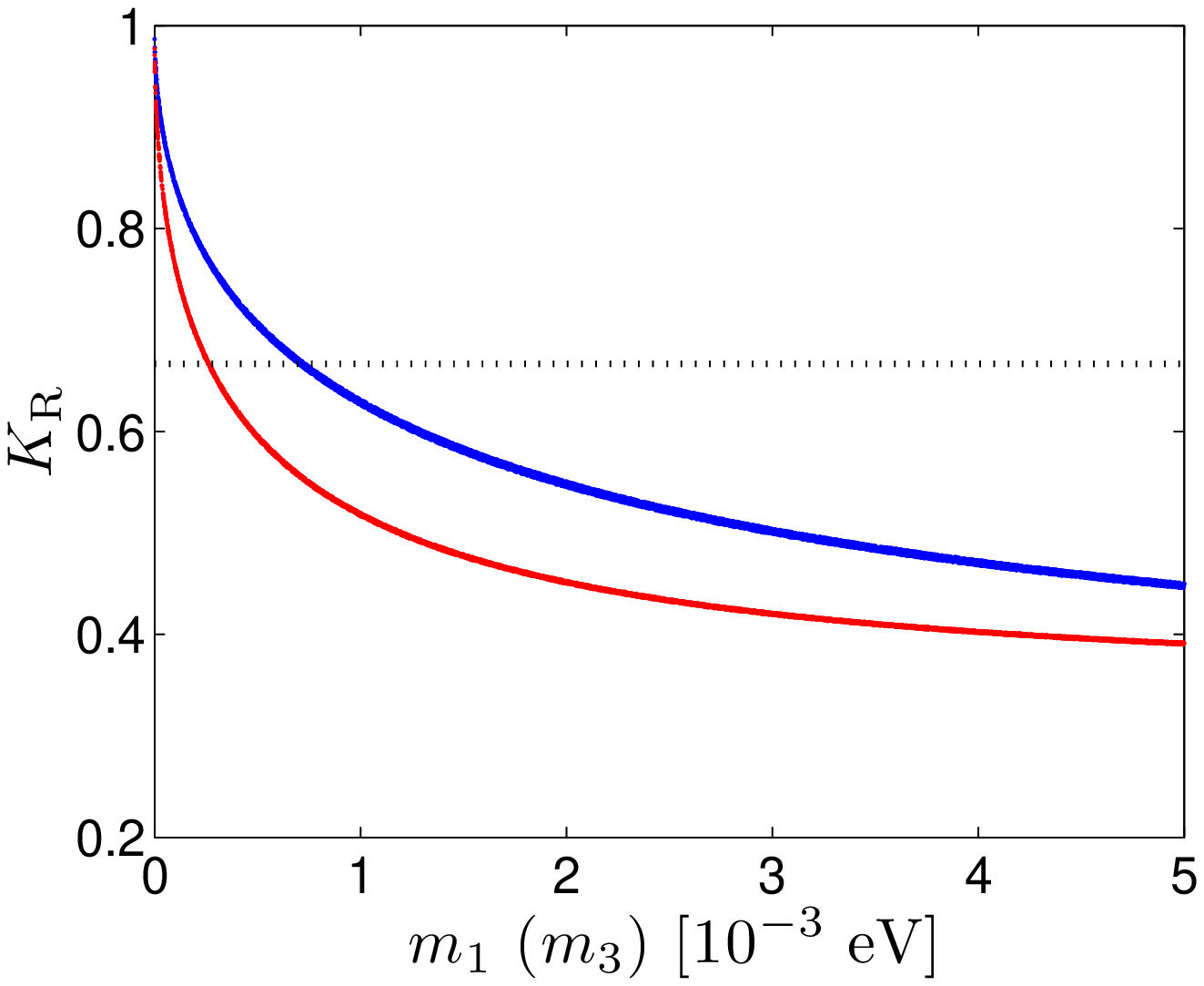}\vspace{0.2cm}
\includegraphics[width=7cm]{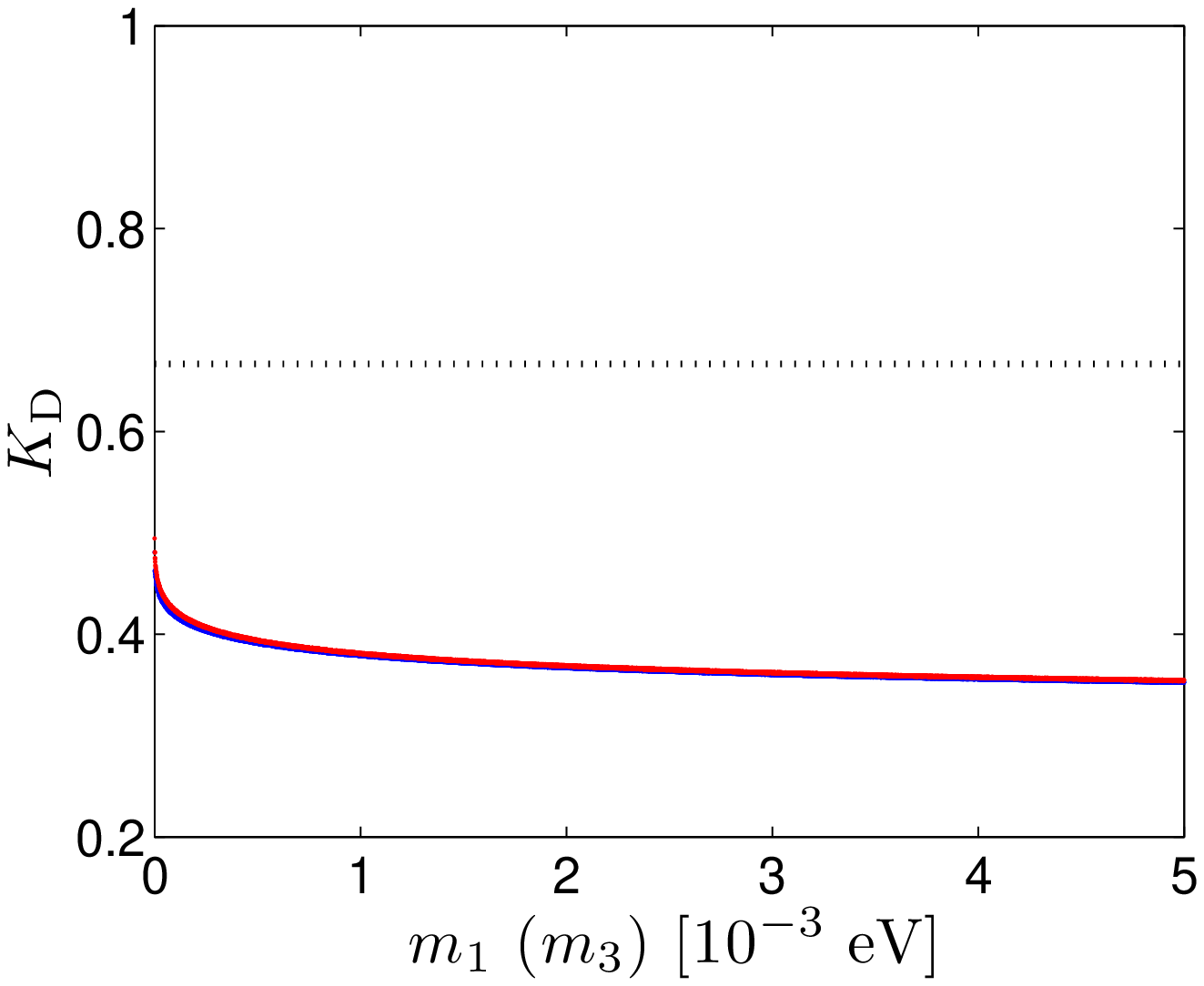}
\vspace{0.cm} \caption{\label{fig:fig1} The dependence of $K_{\rm
R}$ (upper plot) and $K_{\rm D}$ (lower plot) on the lightest
neutrino mass in case I (case II). The red lines correspond
to the normal mass hierarchy, while the blue lines to the inverted
hierarchy. The dotted line corresponds to the exact empirical mass
relation $K_{\rm R} (K_{\rm D})=2/3$.} \vspace{-0.3cm}
\end{center}
\end{figure}
In addition, the dotted line indicates the exact empirical relation,
i.e., $K_{\rm R}=2/3$. One directly finds that the empirical
relation in right-handed neutrino masses can be achieved for
$m_1\simeq 2.5 \times 10^{-4}~{\rm eV}$ in the normal hierarchy case
and $m_3\simeq 8\times 10^{-4}~{\rm eV}$ in the inverted hierarchy
case. These numerical results are in good agreement with the analytical
results.

\item Case II: The empirical relation exists in
$M_{\rm D}$ whereas $M_{\rm R}$ is an identity matrix, i.e.,
$M_i = M_0$. According to Eq.~\eqref{eq:mass}, one has $D_i \sim
\sqrt{m_i}$ indicating that the hierarchy of $D_i$ is milder than
the light neutrino masses. Therefore, the empirical relation cannot
hold, and this case is then ruled out.

In the lower plot of Fig.~\ref{fig:fig1}, we show the value of
$K_{\rm D}$ with respect to the lightest neutrino mass in case
II. The maximum value of $K_{\rm D}$ cannot exceed 0.5, indicating
that case II is incompatible with the experimental data.

\item Case III: Both $M_{\rm D}$ and $M_{\rm R}$ fulfill the empirical
relation. In this case, we have in total six free parameters in
$M_{\rm D}$ and $M_{\rm R}$, out of which five can be fixed by using
Eqs.~\eqref{eq:koide-relation} and \eqref{eq:mass}. One can freely
choose one of the parameters, e.g., $M_1$, since the absolute mass
scale of right-handed neutrinos cannot be determined from the
empirical relation. The light neutrino masses are constrained by the
relation, and we will show in what follows that a nearly degenerate
spectrum is unfavorable.

In our numerical analysis of case III, we fix the mass of one
right-handed neutrino, namely, we take $M_1 =10^9~{\rm GeV}$ in the
normal hierarchy case and $M_3 =10^9~{\rm GeV}$ in the inverted
hierarchy case. Then, the other two right-handed neutrino masses can
be determined from the empirical relation and Eq.~\eqref{eq:mass}
for a given set of light neutrino masses. The allowed ratios between
right-handed neutrino masses are shown in Fig.~\ref{fig:fig2} for
both the normal and inverted hierarchies.
\begin{figure}[t]
\begin{center}\vspace{0.cm}
\includegraphics[width=7cm]{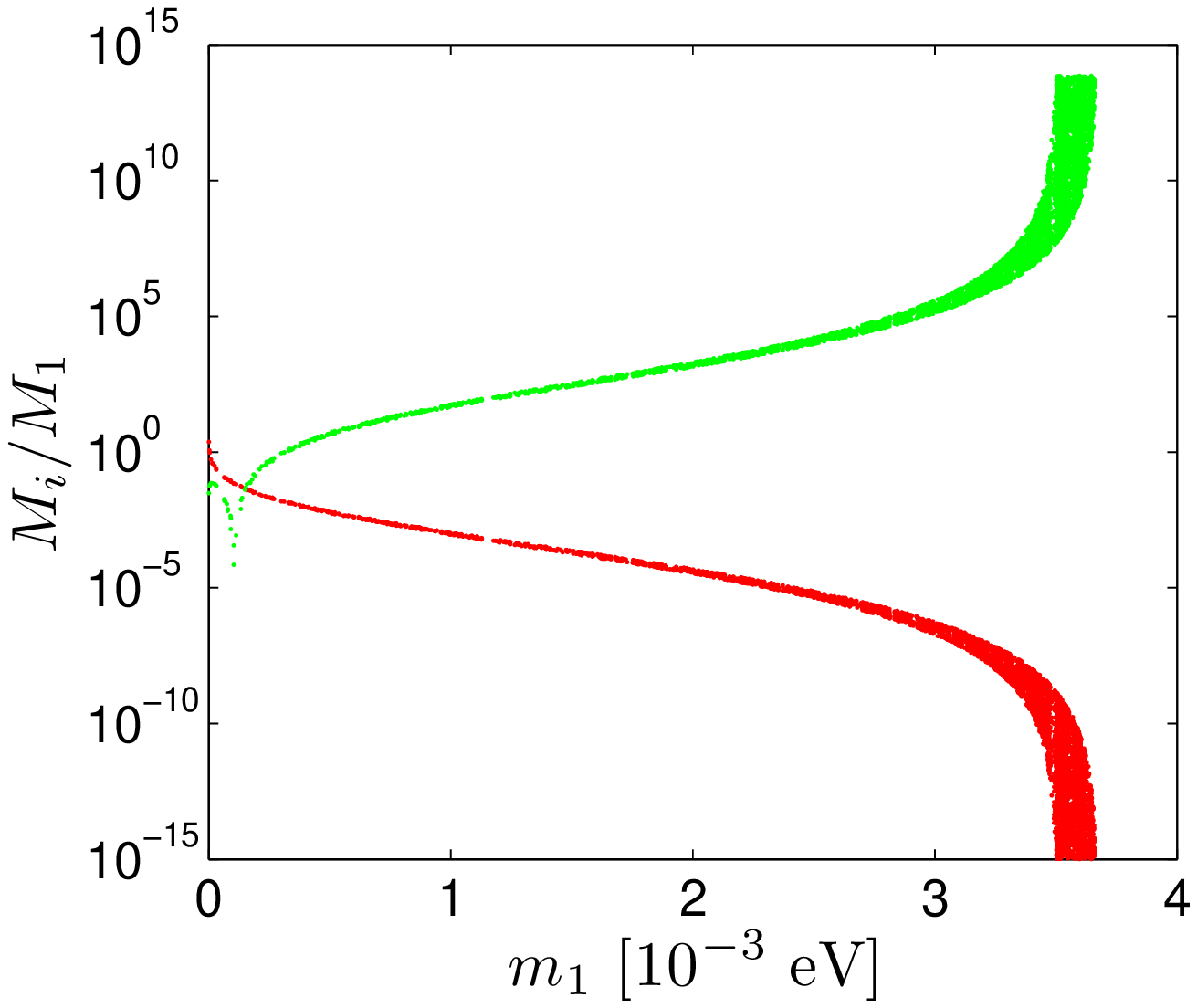}\vspace{0.1cm}
\includegraphics[width=7cm]{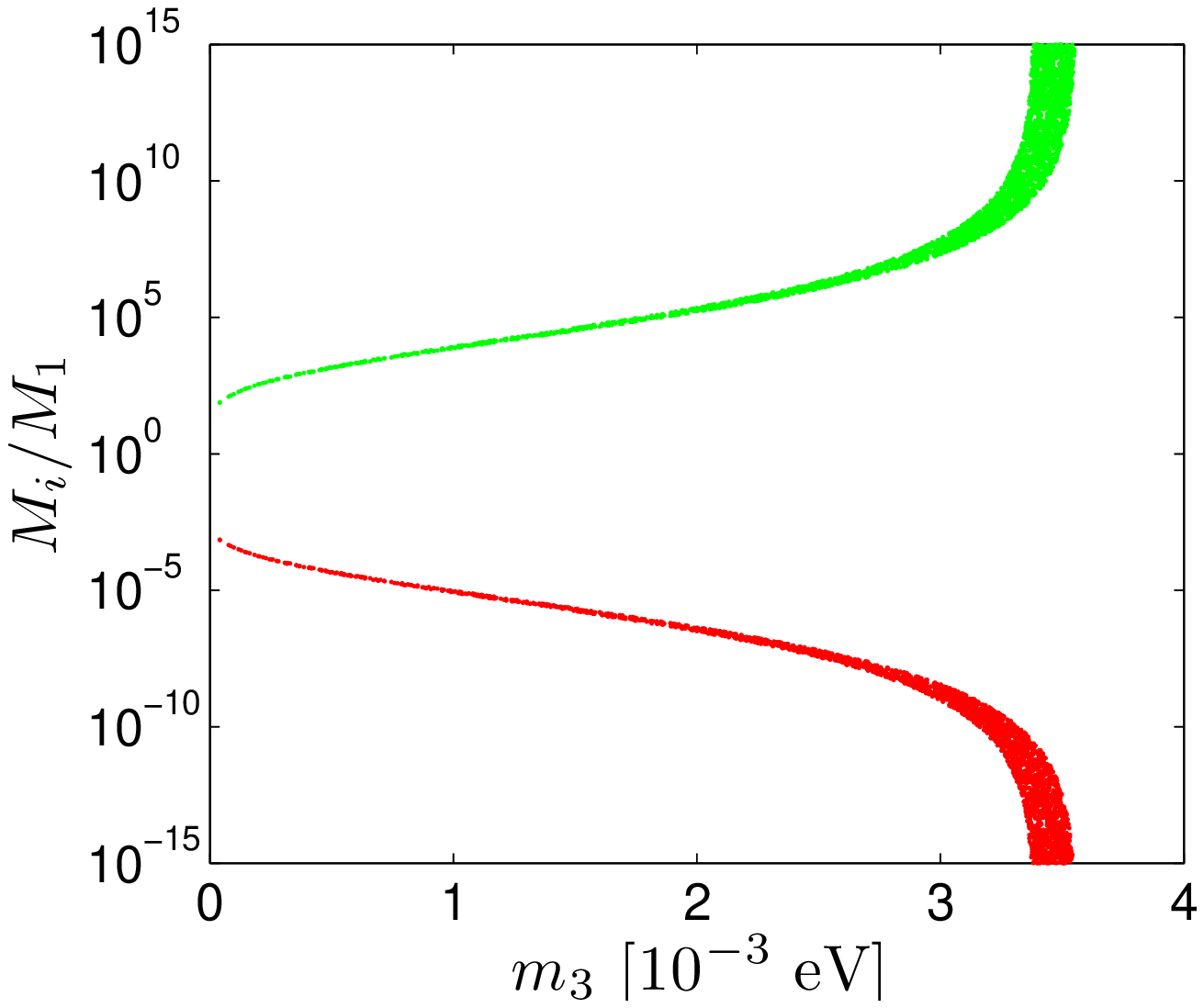}
\vspace{0.15cm} \caption{\label{fig:fig2} Ratios of heavy
right-handed neutrino masses in case III for normal hierarchy (upper
plot) and inverted hierarchy (lower plot). The red and green lines
correspond to the allowed ranges of $M_2/M_1$ and $M_3/M_1$,
respectively.} \vspace{-0.3cm}
\end{center}
\end{figure}
One reads from the plots that there exist stringent upper bounds on
the light neutrino masses, i.e., the mass of the lightest neutrino
cannot exceed $4\times10^{-3}~{\rm eV}$. Furthermore, $M_2$ lies
below $M_1$ and $M_3$ in the inverted hierarchy case, whereas, in
the normal hierarchy case, there exists a flip between $M_2$ and
$M_3$ for $m_1 \sim 1.5\times 10^{-4}~{\rm eV}$. A simple estimate
shows that, at the flip point, $M_2/M_1 =M_3/M_1 \simeq 0.015$. We
stress that the results for this case do not depend on the chosen
value of $M_1$.

\item Case IV: The eigenvalues of $M_{\rm D}$ are the same as the
charged-lepton masses $m_e$, $m_\mu$ and $m_\tau$, respectively,
whereas $M_{\rm R}$ remains unconstrained. Such a special structure
of $M_{\rm D}$ could arise in some grand unification
theories. In this case, the right-handed neutrino masses can be
directly obtained according to Eq.~\eqref{eq:mass} once the light
neutrino masses are chosen.

The predicted right-handed neutrino masses in case IV are
illustrated in Fig.~\ref{fig:fig3}.
\begin{figure}[t]
\begin{center}\vspace{0.cm}
\includegraphics[width=7cm]{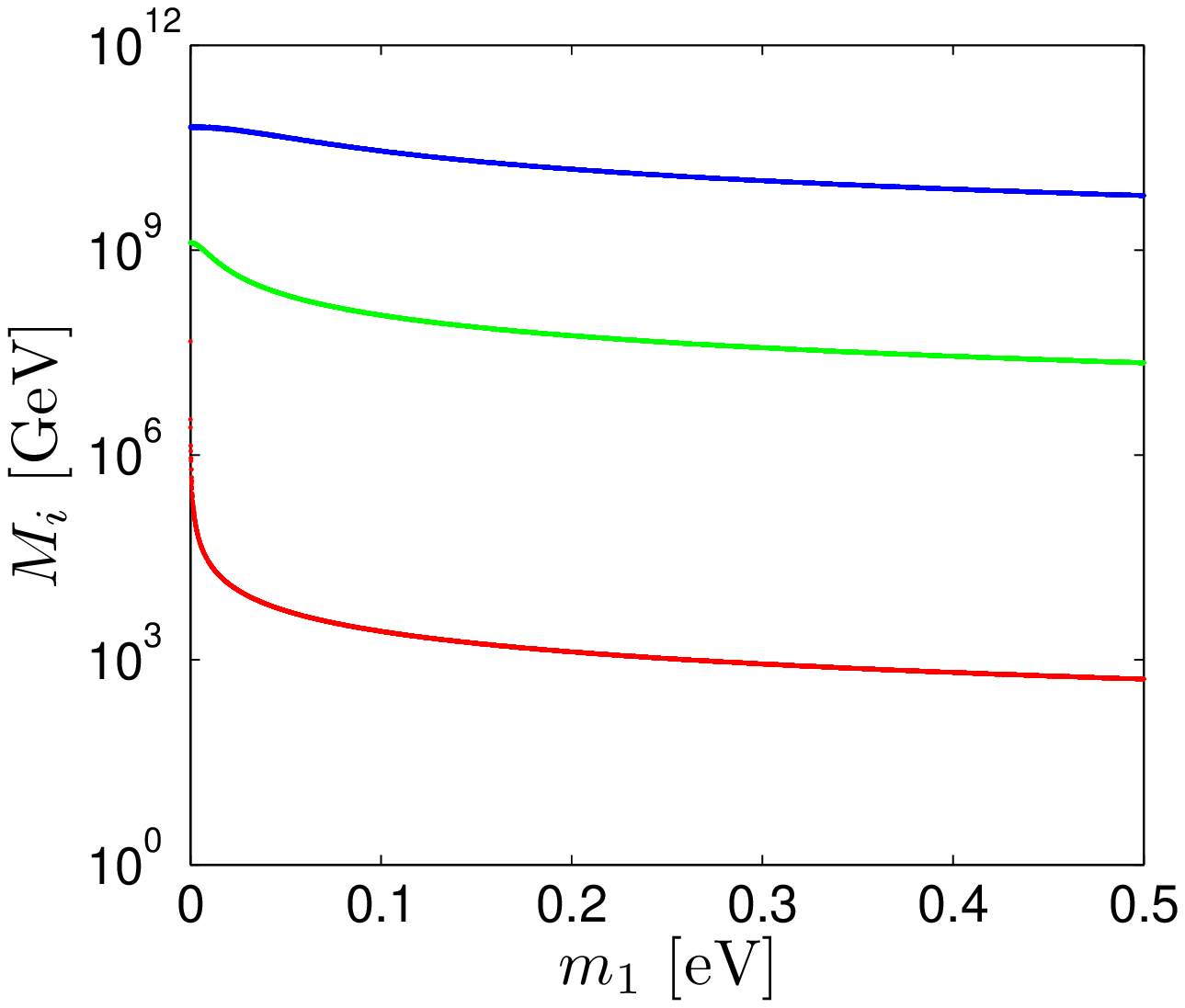}\vspace{0.3cm}
\includegraphics[width=7cm]{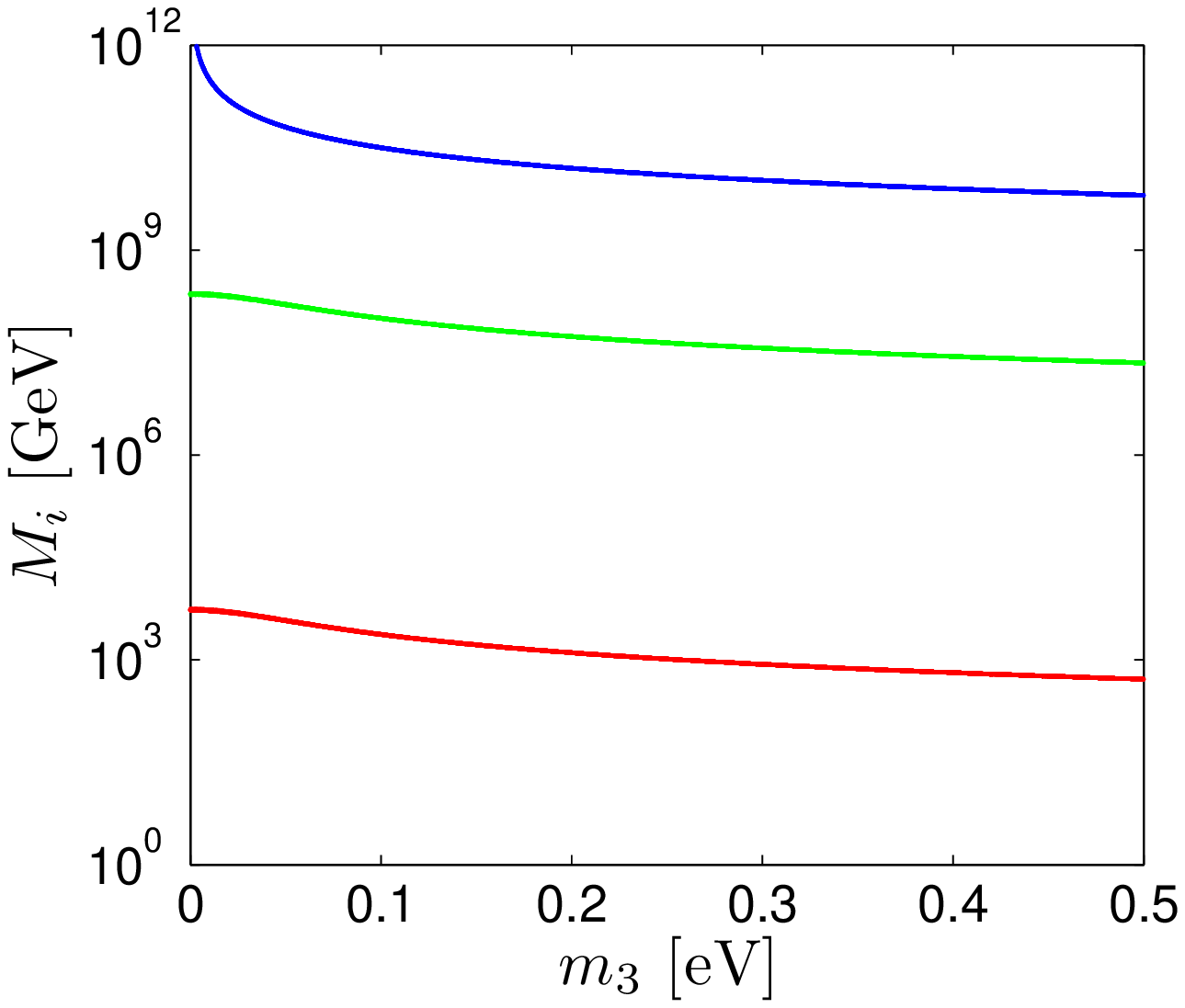}
\vspace{0.cm} \caption{\label{fig:fig3} Ratios of heavy right-handed
neutrino masses in case IV for normal hierarchy (upper plot) and
inverted hierarchy (lower plot). The red and green lines correspond
to the allow ranges of $M_2/M_1$ and $M_3/M_1$, respectively.}
\vspace{-0.3cm}
\end{center}
\end{figure}
In both the normal and inverted hierarchy cases, the right-handed
neutrinos possess a hierarchical spectrum with $M_1 \ll M_2 \ll
M_3$. The mass of lightest right-handed neutrino $M_1$ could be
located around the TeV scale well within the scope of current
colliders, while the heaviest right-handed neutrino mass $M_3$ is
greater than $10^{9} ~{\rm GeV}$.

\end{itemize}

\section{variation of the empirical relation}
\label{sec:variation}

As shown above, the pure light neutrino masses cannot fulfill the
empirical mass relation. However, due to the Majorana nature of
light neutrinos, the deviation of light neutrino masses from the
empirical relations might be viewed as the effects of non-vanishing
Majorana phases. In this sense, it is worth investigating a
variation of the empirical relation, in which Majorana CP-violating
phases are included, i.e.,
\begin{eqnarray}\label{eq:K-phase}
\tilde K_{\nu} = \left|\frac{m_1+m_2 {\rm e}^{i\phi_1}+m_3 {\rm
e}^{i\phi_2}}{\left(\sqrt{m_1}+\sqrt{m_2 }{\rm
e}^{i\frac{\phi_1}{2}}+\sqrt{m_3 }{\rm
e}^{i\frac{\phi_2}{2}}\right)^2} \right|\, .
\end{eqnarray}
Allowing $\phi_1$ and $\phi_2$ to vary between 0 and $2\pi$, $\tilde
K_{\nu} = 2/3$ can be easily achieved, since there are two more free
parameters entering the relation. In particular, $\tilde K_{\nu}$
could be larger than 1 or close to zero if there exists strong
cancellation in the denominator or nominator, which is quite
relevant in the case of a nearly degenerate light neutrino mass
spectrum.

\begin{figure}[t]
\begin{center}\vspace{0.cm}
\includegraphics[width=7.2cm]{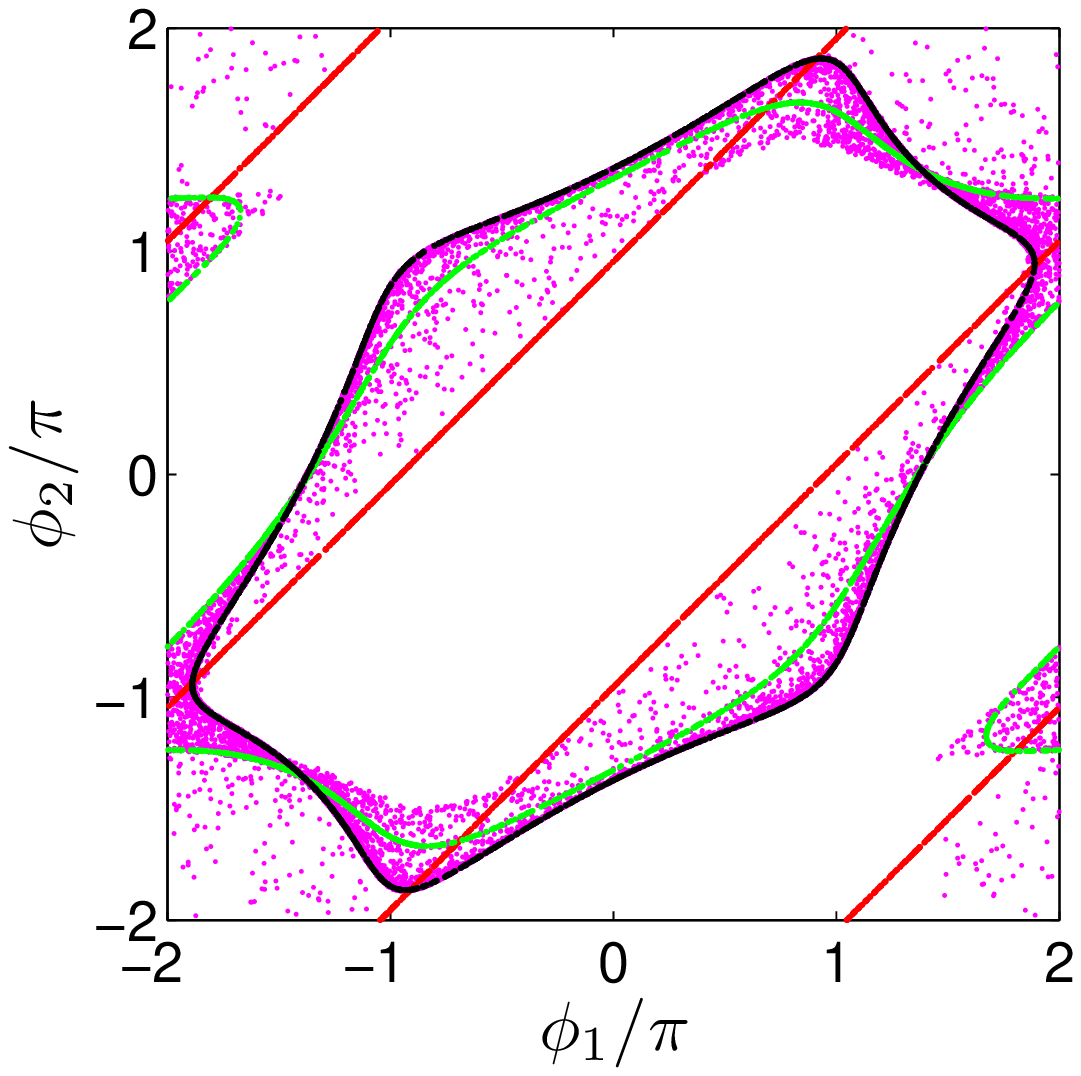}\vspace{0.1cm}
\includegraphics[width=7.2cm]{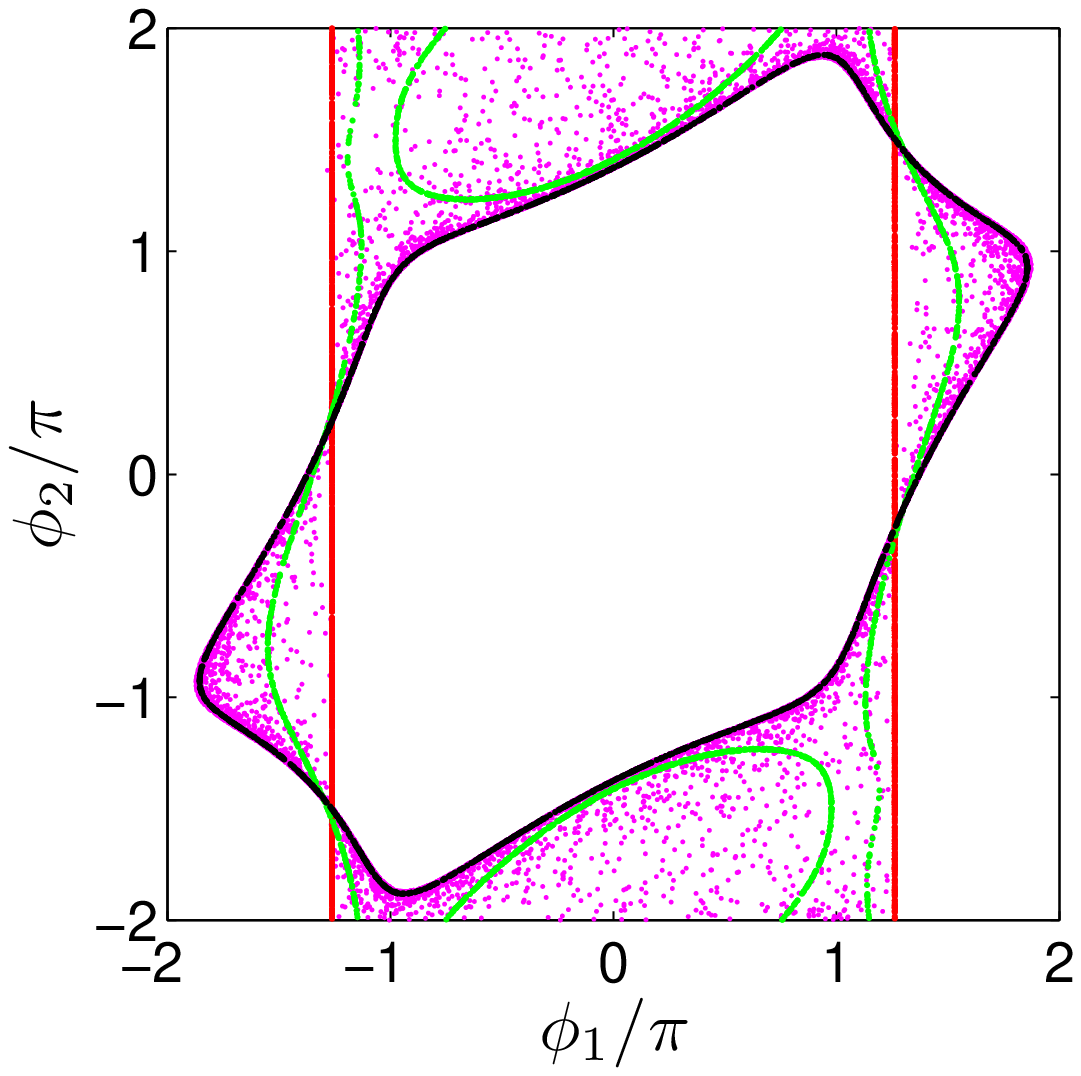}
\vspace{-0.1cm} \caption{\label{fig:fig4} The allowed parameter
spaces of $\phi_1$ and $\phi_2$ for normal hierarchy (upper plot)
and inverted hierarchy (lower plot). The red, green, blue and black
lines in the plot correspond to the lightest mass $m_1
(m_3)=0,~0.02~{\rm eV}$, and $0.2~{\rm eV}$, respectively.}
\vspace{-0.3cm}
\end{center}
\end{figure}

The allowed parameter spaces of $\phi_1$ and $\phi_2$ required for
$\tilde K_{\nu}=2/3$ are shown in Fig.~\ref{fig:fig4}.
We see that the two Majorana phases are constrained by the empirical
relation, e.g., $\phi_1=\phi_2=0$ is unfavored in both hierarchies.
In the nearly degenerate limit, the correlation between $\phi_1$ and
$\phi_2$ is basically the same for normal and inverted ordering,
since in the limit $m_1\simeq m_2 \simeq m_3$,
\begin{eqnarray}\label{eq:K-deg}
\tilde K_{\nu} \simeq \left| \frac{1+ {\rm e}^{i\phi_1}+{\rm
e}^{i\phi_2}}{\left(1+ {\rm e}^{i\frac{\phi_1}{2}}+{\rm
e}^{i\frac{\phi_2}{2}}\right)^2} \right|\, ,
\end{eqnarray}
holds for both cases.

In case of a normal and hierarchical mass spectrum, $\tilde K_\nu$
turns out to be only sensitive to the phase difference
$\phi=\phi_2-\phi_1$, and according to the plot $\phi$ should be
very close to $-\pi$. This can be understood from
Eq.~\eqref{eq:K-phase} which, in the limit $m_1\sim 0$, can be reduced to
\begin{eqnarray}\label{eq:K-nor}
\tilde K_{\nu} \simeq \left| \frac{1+r^2 {\rm e}^{i\phi}}{1+2r{\rm
e}^{i\frac{\phi}{2}}+r^2 {\rm e}^{i\phi}} \right|\, ,
\end{eqnarray}
where $r=\sqrt{m_2/m_3} \simeq 0.4$, and $\phi=\phi_1-\phi_2$.
Taking $\phi=-\pi$, one can estimate that $\tilde K_\nu \simeq 0.7$,
roughly in agreement with the empirical relation.

An interesting result is found for the inverted hierarchy
($m_2 \simeq m_1 \gg m_3$). One has no dependence on $\phi_2$, and
finds
\begin{eqnarray}\label{eq:K-inv}
\tilde K_{\nu} \simeq \left| \frac{1+ {\rm e}^{i\phi_1}}{1+2{\rm
e}^{i\frac{\phi_1}{2}}+{\rm e}^{i\phi_1}} \right|\, .
\end{eqnarray}
This expression gives $\tilde K_{\nu} = \frac 23$ for a phase which
corresponds to $\sin^2\phi_1/2 = \frac{21}{25}$.

\begin{figure}[t]
\begin{center}\vspace{0.cm}
\includegraphics[width=7cm]{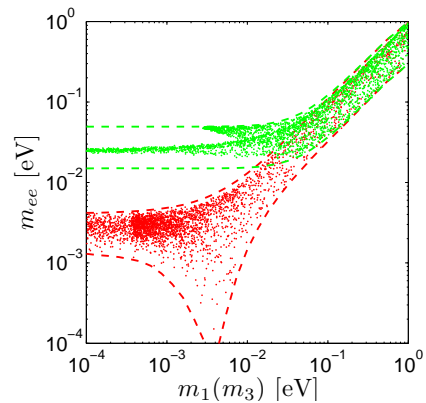}
\vspace{0.cm} \caption{\label{fig:fig5} The allowed range of the
effective mass $m_{ee}$ as a function of the lightest neutrino mass
in the normal hierarchy (red line) and inverted hierarchy (green
line) at 1$\sigma$ C.L.~with $\tilde K_{\nu}=2/3$ being satisfied.
The dashed lines correspond to the parameter range without assuming
$\tilde K_{\nu}=2/3$.} \vspace{-0.3cm}
\end{center}
\end{figure}

Since the Majorana nature of neutrinos is aimed to be revealed in
future neutrinoless double decay experiments, there is a need to
address some comments on the connection between the empirical mass
relation and the neutrinoless double beta decay process, in which
the decay amplitude is proportional to
\begin{eqnarray}\label{eq:meff}
m_{ee} & = & \left| \sum {V}^2_{ei} m_i\right| \\ & = &
\left||V_{e1}|^2 m_1 + |V_{e2}|^2 {\rm e}^{i \phi_1} m_2 +
|V_{e3}|^2 {\rm e}^{i \phi_2} m_3 \right| \nonumber  \, .
\end{eqnarray}
Here the leptonic flavor mixing matrix $V$ is given by
$V_{e1}=\cos\theta_{12}\cos\theta_{13}$,
$V_{e2}=\sin\theta_{12}\cos\theta_{13}$, and
$V_{e3}=\sin\theta_{13}$, with $\theta_{ij}$ being the lepton mixing
angles. We illustrate in Fig.~\ref{fig:fig5} the constraints on
$m_{ee}$ when $\tilde K_{\nu}=2/3$ is satisfied.
In the normal hierarchy case, the predicted $m_{ee}$ almost saturates
all of the experimentally allowed range for $m_1 \lesssim 1$ eV.
However, in the inverted hierarchy case, the allowed range of
$m_{ee}$ shrinks a lot if $m_3 \lesssim 0.01~{\rm eV}$. This is a
consequence of Eq.~(\ref{eq:K-inv}), and the value $\sin^2\phi_1/2 =
\frac{21}{25}$ following from the requirement $\tilde K_\nu = \frac
23$. Inserting this phase in the expression for the effective mass,
whose general value is
\begin{eqnarray}
m_{ee} & = & \nonumber
\cos^2 \theta_{13} \sqrt{\dma} \sqrt{1 - \sin^2 2 \theta_{12} \,
\sin^2 \phi_1/2} \,  ,
\end{eqnarray}
which gives the remarkable result
\begin{eqnarray}
m_{ee} & = & \frac 15 \cos^2 \theta_{13} \sqrt{\dma} \sqrt{25 - 21 \sin^2 2
\theta_{12}} \, .
\end{eqnarray}
This fixes the effective mass to about 0.025 eV,
with little dependence on the oscillation parameters. One should
compare this value with the general lower and upper limits of 0.017 and 0.05 eV,
respectively.

\section{conclusion}\label{sec:summary}

In this work we have added several new points regarding the
properties of an empirical mass relation (Koide relation), which
frequently is discussed in the literature.

We first extended the empirical mass relation of charged leptons to
the other fermion sectors. In particular, we have noted that there
could exist an universal empirical relation in the quark sectors
once the quarks are classified by their mass scales instead of their
electric charge or isospin. We then noted that if light neutrinos
acquire their mass via the seesaw mechanism, it is not surprising
that they fail to obey the empirical relation. Instead, we applied
the relation to the Dirac and/or heavy Majorana mass matrix. We
illustrated the consequences for neutrino masses in a simplified
seesaw framework by assuming all lepton mixing to stem from the
charged lepton sector, and analyzed four typical scenarios realizing
the empirical relation in different ways. Furthermore, we
generalized the empirical mass relation with Majorana phases being
included and found that in case of an inverted hierarchy mass
spectrum, the effective mass is strongly constrained. The relevant
Majorana phase is fixed and a value of about $m_{ee} \simeq
0.025~{\rm eV}$ exists, being a factor of 1.5 larger than the
general lower bound.

Let us remark here that in a more general seesaw framework, there
will be much more free parameters entering the expression of $M_{\rm
D}$, and there are in principle enough degrees of freedom to
reproduce the empirical relation in both $M_{\rm D}$ and $M_{\rm R}$
for any choice of light neutrino masses. In addition, thermal
leptogenesis cannot work in our simplified framework even if flavor
effects are taken into account. Note that the empirical mass
relation discussed in this work could be re-scaled up to a higher
and more fundamental scale, i.e., the grand unification scale, by
using RG equations. However, as shown in Ref.~\cite{Xing:2006vk},
the fermion mass ratios tend to be very stable against radiative
corrections even in the supersymmetric model with a large
$\tan\beta$. If there are intermediate scales in the RG evolution,
the running fermion masses could be modified by the threshold
effects, in particular for neutrinos. Such a study could be done in
a model-dependent manner and lies beyond the scope of current work.

\begin{acknowledgments}
This work was supported by the ERC under the Starting Grant MANITOP
and the Deutsche Forschungsgemeinschaft in the Transregio 27
``Neutrinos and beyond -- weakly interacting particles in physics,
astrophysics and cosmology''.
\end{acknowledgments}

\bibliography{bib}

\begin{thebibliography}{23}
\expandafter\ifx\csname natexlab\endcsname\relax\def\natexlab#1{#1}\fi
\expandafter\ifx\csname bibnamefont\endcsname\relax
  \def\bibnamefont#1{#1}\fi
\expandafter\ifx\csname bibfnamefont\endcsname\relax
  \def\bibfnamefont#1{#1}\fi
\expandafter\ifx\csname citenamefont\endcsname\relax
  \def\citenamefont#1{#1}\fi
\expandafter\ifx\csname url\endcsname\relax
  \def\url#1{\texttt{#1}}\fi
\expandafter\ifx\csname urlprefix\endcsname\relax\def\urlprefix{URL }\fi
\providecommand{\bibinfo}[2]{#2}
\providecommand{\eprint}[2][]{\url{#2}}

\bibitem[{\citenamefont{Koide}(1982)}]{Koide:1982si}
\bibinfo{author}{\bibfnamefont{Y.}~\bibnamefont{Koide}},
  \bibinfo{journal}{Nuovo Cim. Lett.} \textbf{\bibinfo{volume}{34}},
  \bibinfo{pages}{201} (\bibinfo{year}{1982}).

\bibitem[{\citenamefont{Koide}(1983)}]{Koide:1983qe}
\bibinfo{author}{\bibfnamefont{Y.}~\bibnamefont{Koide}},
  \bibinfo{journal}{Phys. Rev.} \textbf{\bibinfo{volume}{D28}},
  \bibinfo{pages}{252} (\bibinfo{year}{1983}).

\bibitem[{\citenamefont{Koide}(1990)}]{Koide:1989jq}
\bibinfo{author}{\bibfnamefont{Y.}~\bibnamefont{Koide}}, \bibinfo{journal}{Mod.
  Phys. Lett.} \textbf{\bibinfo{volume}{A5}}, \bibinfo{pages}{2319}
  (\bibinfo{year}{1990}).

\bibitem[{\citenamefont{Foot}(1994)}]{Foot:1994yn}
\bibinfo{author}{\bibfnamefont{R.}~\bibnamefont{Foot}} (\bibinfo{year}{1994}),
  \eprint{hep-ph/9402242}.

\bibitem[{\citenamefont{Koide and Tanimoto}(1996)}]{Koide:1995xk}
\bibinfo{author}{\bibfnamefont{Y.}~\bibnamefont{Koide}} \bibnamefont{and}
  \bibinfo{author}{\bibfnamefont{M.}~\bibnamefont{Tanimoto}},
  \bibinfo{journal}{Z. Phys.} \textbf{\bibinfo{volume}{C72}},
  \bibinfo{pages}{333} (\bibinfo{year}{1996}), \eprint{hep-ph/9505333}.

\bibitem[{\citenamefont{Koide and Fusaoka}(1997)}]{Koide:1996yf}
\bibinfo{author}{\bibfnamefont{Y.}~\bibnamefont{Koide}} \bibnamefont{and}
  \bibinfo{author}{\bibfnamefont{H.}~\bibnamefont{Fusaoka}},
  \bibinfo{journal}{Prog. Theor. Phys.} \textbf{\bibinfo{volume}{97}},
  \bibinfo{pages}{459} (\bibinfo{year}{1997}), \eprint{hep-ph/9612322}.

\bibitem[{\citenamefont{Rivero and Gsponer}(2005)}]{Rivero:2005vj}
\bibinfo{author}{\bibfnamefont{A.}~\bibnamefont{Rivero}} \bibnamefont{and}
  \bibinfo{author}{\bibfnamefont{A.}~\bibnamefont{Gsponer}}
  (\bibinfo{year}{2005}), \eprint{hep-ph/0505220}.

\bibitem[{\citenamefont{Krolikowski}(2005)}]{Krolikowski:2005kr}
\bibinfo{author}{\bibfnamefont{W.}~\bibnamefont{Krolikowski}}
  (\bibinfo{year}{2005}), \eprint{hep-ph/0508039}.

\bibitem[{\citenamefont{Gerard et~al.}(2006)\citenamefont{Gerard, Goffinet, and
  Herquet}}]{Gerard:2005ad}
\bibinfo{author}{\bibfnamefont{J.~M.} \bibnamefont{Gerard}},
  \bibinfo{author}{\bibfnamefont{F.}~\bibnamefont{Goffinet}}, \bibnamefont{and}
  \bibinfo{author}{\bibfnamefont{M.}~\bibnamefont{Herquet}},
  \bibinfo{journal}{Phys. Lett.} \textbf{\bibinfo{volume}{B633}},
  \bibinfo{pages}{563} (\bibinfo{year}{2006}), \eprint{hep-ph/0510289}.

\bibitem[{\citenamefont{Koide}(2009)}]{Koide:2008tr}
\bibinfo{author}{\bibfnamefont{Y.}~\bibnamefont{Koide}},
  \bibinfo{journal}{Phys. Rev.} \textbf{\bibinfo{volume}{D79}},
  \bibinfo{pages}{033009} (\bibinfo{year}{2009}), \eprint{0811.3470}.

\bibitem[{\citenamefont{Sumino}(2009{\natexlab{a}})}]{Sumino:2008hu}
\bibinfo{author}{\bibfnamefont{Y.}~\bibnamefont{Sumino}},
  \bibinfo{journal}{Phys. Lett.} \textbf{\bibinfo{volume}{B671}},
  \bibinfo{pages}{477} (\bibinfo{year}{2009}{\natexlab{a}}),
  \eprint{0812.2090}.

\bibitem[{\citenamefont{Sumino}(2009{\natexlab{b}})}]{Sumino:2008hy}
\bibinfo{author}{\bibfnamefont{Y.}~\bibnamefont{Sumino}},
  \bibinfo{journal}{JHEP} \textbf{\bibinfo{volume}{05}}, \bibinfo{pages}{075}
  (\bibinfo{year}{2009}{\natexlab{b}}), \eprint{0812.2103}.

\bibitem[{\citenamefont{Sumino}(2009{\natexlab{c}})}]{Sumino:2009bt}
\bibinfo{author}{\bibfnamefont{Y.}~\bibnamefont{Sumino}}
  (\bibinfo{year}{2009}{\natexlab{c}}), \eprint{0903.3640}.

\bibitem[{\citenamefont{Koide}(2010)}]{Koide:2010np}
\bibinfo{author}{\bibfnamefont{Y.}~\bibnamefont{Koide}},
  \bibinfo{journal}{Phys. Lett.} \textbf{\bibinfo{volume}{B687}},
  \bibinfo{pages}{219} (\bibinfo{year}{2010}), \eprint{1001.4877}.

\bibitem[{\citenamefont{Li and Ma}(2006)}]{Li:2006et}
\bibinfo{author}{\bibfnamefont{N.}~\bibnamefont{Li}} \bibnamefont{and}
  \bibinfo{author}{\bibfnamefont{B.-Q.} \bibnamefont{Ma}},
  \bibinfo{journal}{Phys. Rev.} \textbf{\bibinfo{volume}{D73}},
  \bibinfo{pages}{013009} (\bibinfo{year}{2006}), \eprint{hep-ph/0601031}.

\bibitem[{\citenamefont{Xing and Zhang}(2006)}]{Xing:2006vk}
\bibinfo{author}{\bibfnamefont{Z.-z.} \bibnamefont{Xing}} \bibnamefont{and}
  \bibinfo{author}{\bibfnamefont{H.}~\bibnamefont{Zhang}},
  \bibinfo{journal}{Phys. Lett.} \textbf{\bibinfo{volume}{B635}},
  \bibinfo{pages}{107} (\bibinfo{year}{2006}), \eprint{hep-ph/0602134}.

\bibitem[{\citenamefont{Nakamura et~al.}(2010)}]{Nakamura:2010zzi}
\bibinfo{author}{\bibfnamefont{K.}~\bibnamefont{Nakamura}} \bibnamefont{et~al.}
  (\bibinfo{collaboration}{Particle Data Group}), \bibinfo{journal}{J. Phys.}
  \textbf{\bibinfo{volume}{G37}}, \bibinfo{pages}{075021}
  (\bibinfo{year}{2010}).

\bibitem[{\citenamefont{Xing et~al.}(2008)\citenamefont{Xing, Zhang, and
  Zhou}}]{Xing:2007fb}
\bibinfo{author}{\bibfnamefont{Z.-z.} \bibnamefont{Xing}},
  \bibinfo{author}{\bibfnamefont{H.}~\bibnamefont{Zhang}}, \bibnamefont{and}
  \bibinfo{author}{\bibfnamefont{S.}~\bibnamefont{Zhou}},
  \bibinfo{journal}{Phys. Rev.} \textbf{\bibinfo{volume}{D77}},
  \bibinfo{pages}{113016} (\bibinfo{year}{2008}), \eprint{0712.1419}.

\bibitem[{\citenamefont{Minkowski}(1977)}]{Minkowski:1977sc}
\bibinfo{author}{\bibfnamefont{P.}~\bibnamefont{Minkowski}},
  \bibinfo{journal}{Phys. Lett.} \textbf{\bibinfo{volume}{B67}},
  \bibinfo{pages}{421} (\bibinfo{year}{1977}).

\bibitem[{\citenamefont{Yanagida}(1979)}]{Yanagida:1979as}
\bibinfo{author}{\bibfnamefont{T.}~\bibnamefont{Yanagida}}, in
  \emph{\bibinfo{booktitle}{Proc. Workshop on the baryon number of the Universe
  and unified theories}}, edited by
  \bibinfo{editor}{\bibfnamefont{O.}~\bibnamefont{Sawada}} \bibnamefont{and}
  \bibinfo{editor}{\bibfnamefont{A.}~\bibnamefont{Sugamoto}}
  (\bibinfo{year}{1979}), p.~\bibinfo{pages}{95}.

\bibitem[{\citenamefont{Gell-Mann et~al.}(1979)\citenamefont{Gell-Mann, Ramond,
  and Slansky}}]{GellMann:1980vs}
\bibinfo{author}{\bibfnamefont{M.}~\bibnamefont{Gell-Mann}},
  \bibinfo{author}{\bibfnamefont{P.}~\bibnamefont{Ramond}}, \bibnamefont{and}
  \bibinfo{author}{\bibfnamefont{R.}~\bibnamefont{Slansky}}, in
  \emph{\bibinfo{booktitle}{Supergravity}}, edited by
  \bibinfo{editor}{\bibfnamefont{P.}~\bibnamefont{van Nieuwenhuizen}}
  \bibnamefont{and} \bibinfo{editor}{\bibfnamefont{D.}~\bibnamefont{Freedman}}
  (\bibinfo{year}{1979}), p. \bibinfo{pages}{315}.

\bibitem[{\citenamefont{Mohapatra and Senjanovic}(1980)}]{Mohapatra:1979ia}
\bibinfo{author}{\bibfnamefont{R.~N.} \bibnamefont{Mohapatra}}
  \bibnamefont{and}
  \bibinfo{author}{\bibfnamefont{G.}~\bibnamefont{Senjanovic}},
  \bibinfo{journal}{Phys. Rev. Lett.} \textbf{\bibinfo{volume}{44}},
  \bibinfo{pages}{912} (\bibinfo{year}{1980}).

\bibitem[{\citenamefont{Schwetz et~al.}(2008)\citenamefont{Schwetz, Tortola,
  and Valle}}]{Schwetz:2008er}
\bibinfo{author}{\bibfnamefont{T.}~\bibnamefont{Schwetz}},
  \bibinfo{author}{\bibfnamefont{M.~A.} \bibnamefont{Tortola}},
  \bibnamefont{and} \bibinfo{author}{\bibfnamefont{J.~W.~F.}
  \bibnamefont{Valle}}, \bibinfo{journal}{New J. Phys.}
  \textbf{\bibinfo{volume}{10}}, \bibinfo{pages}{113011}
  (\bibinfo{year}{2008}), \eprint{0808.2016}.

\end{thebibliography}

\end{document}